\documentclass[usenatbib]{mn2e}
\usepackage{natbib}  
\usepackage{epsfig}

\catcode`\@=11
\def\gta{\ifmmode{\,\mathrel{\mathpalette\@versim>\,}}
    \else{$\,\mathrel{\mathpalette\@versim>}\,$}\fi}
\def\lta{\ifmmode{\,\mathrel{\mathpalette\@versim<\,}}
    \else{$\,\mathrel{\mathpalette\@versim<}\,$}\fi}
\def\@versim#1#2{\lower 2.9truept \vbox{\baselineskip 0pt \lineskip
    0.5truept \ialign{$\m@th#1\hfil##\hfil$\crcr#2\crcr\sim\crcr}}}
\catcode`\@=12  
\renewcommand{\[}{\begin{equation}}
\renewcommand{\]}{\end{equation}}
 
\def\cL{{\cal L}}
\def\ex#1{\la#1\ra}

\def\p{\partial}
\def\mcmc{{\sc mcmc}}
\font\gkvecten=cmmib10
\font\gkvecseven=cmmib7
\let\boldgrk=\gkvecten
\let\boldgrksc=\gkvecseven

\def\gkthing#1{{\mathchoice%
	{\hbox{{\boldgrk\char#1}}}
	{\hbox{{\boldgrk\char#1}}}
	{\hbox{{\boldgrksc\char#1}}}
	{\hbox{{\boldgrksc\char#1}}}}}

\def\vtheta{\gkthing{18}}

{\newif\ifnotend
\notendtrue
\def\veclist{ABCDEFGHIJKLMNOPQRSTUVWXYZabcdefghijklmnopqrstuvwxyz.}
\def\top#1#2.{#1}
\def\tail#1#2.{#2.}
\loop\expandafter\xdef\csname v\expandafter\top\veclist\endcsname%
{{\noexpand\bf\expandafter\top\veclist}}
\edef\veclist{\expandafter\tail\veclist}
\if\veclist.\notendfalse\fi\ifnotend\repeat}
 
\def\df{{\sc df}}\def\aa{{\sc aa}}

\def\fracj#1#2{{\textstyle{#1\over#2}}}

\def\ex#1{\left\langle#1\right\rangle}
\def\Rc{R_{\rm c}}

\def\kms{\,{\rm km}\,{\rm s}^{-1}}

\def\K{\,{\rm K}} 
\def\pc{\,{\rm pc}}
\def\kpc{\,{\rm kpc}}

\def\e{{\rm e}}
\def\d{{\rm d}}
\def\msun{\,{\rm M}_\odot}

\def\figref#1{Fig.~\ref{#1}}
\newcommand{\beq}{\begin{equation}}
\newcommand{\eeq}{\end{equation}}

\title[Models of our Galaxy II]
{Models  of our Galaxy II}

\author[J. Binney and P. McMillan]{James
Binney$^1$\thanks{E-mail:
binney@thphys.ox.ac.uk} and Paul McMillan$^1$\\
$^{1}$ Rudolf Peierls Centre for Theoretical Physics, Keble Road, Oxford OX1 3NP, UK\\
}
\begin{document}

\date{Draft, September 8, 2010}

\pagerange{\pageref{firstpage}--\pageref{lastpage}} \pubyear{2009}

\maketitle

\label{firstpage}

\begin{abstract}

Stars near the Sun oscillate both horizontally and vertically. In Paper I it
was assumed that the coupling between these motions can be modelled by
determining the horizontal motion without reference to the vertical motion,
and recovering the coupling between the motions by assuming that the vertical
action is adiabatically conserved as the star oscillates horizontally. Here
we show that, although the assumption of adiabatic invariance works well,
more accurate results can be obtained by taking the vertical action into
account when calculating the horizontal motion.  We use orbital tori to
present a simple but fairly realistic model of the Galaxy's discs in which
the motion of stars is handled rigorously, without decomposing it into
horizontal and vertical components. We examine the ability of the adiabatic
approximation to calculate the model's observables, and find that it performs
perfectly in the plane, but errs slightly away from the plane. When the new
correction to the adiabatic approximation is used, the density,
mean-streaming velocity and velocity dispersions  are in
error by less than 10 per cent for distances up to $2.5\kpc$ from the Sun.
The torus-based model reveals that at locations above the plane the long axis
of the velocity ellipsoid points almost to the Galactic centre,
even though the model potential is significantly flattened.  This result
contradicts the widespread belief that the shape of the Galaxy's potential
can be strongly constrained by the orientation of velocity ellipsoid near the
Sun. An analysis of individual orbits reveals that in
a general potential the orientation of the velocity ellipsoid depends on the
structure of the model's distribution function as much as on its
gravitational potential, contrary to what is the case for St\"ackel
potentials. We argue that the adiabatic approximation will provide a valuable
complement to torus-based models in the interpretation of current surveys of
the Galaxy.

\end{abstract}


\section{Introduction}
Study of the structure of the Milky Way Galaxy is a major theme of
contemporary astronomy. Our Galaxy is typical of the galaxies that dominate
the current cosmic star-formation rate, so understanding it is of more than
parochial interest. We believe that most of its mass is contributed by
elementary particles that have yet to be directly detected, but we have only
weak constraints on the spatial density and kinematics of these particles --
we urgently need stronger constraints on them. The Cold-Dark-Matter (CDM)
cosmogony provides a very persuasive picture of how a galaxy such as ours
formed, and we need to know how accurately this theory predicts the structure
of the Galaxy.

In view of these considerations, large resources have been invested over the
last decade in massive surveys of the stellar content of the Galaxy.  The
rate at which data from this observational effort becomes available will
increase at least through 2020. Models of the Galaxy will surely play a key
role in extracting science from the data, because the Galaxy is a very
complex object and every survey is subject to powerful observational biases.
Consequently it is extremely hard to proceed in a model-independent way from
observational data to physical understanding. We are likely to achieve the
desired physical understanding by comparing observational data with
the predictions of models. 

It is useful to distinguish between  kinematic and dynamic models. A
kinematic model specifies the spatial density of stars and their kinematics
at each point without asking whether a gravitational potential exists in which
the given density distribution and kinematics constitute a steady state.
\cite{BahcallS} pioneered kinematic models, and recent versions include
Galaxia \citep{Sharma10}. The science of constructing
dynamical models is still in its infancy. The Besan\c con model
\citep{Robin03} has a dynamical element to it in that in it the disc's density
profile perpendicular to the plane is dynamically consistent with the
corresponding  run of dispersion of velocities perpendicular to the plane.
\cite{SB09} and \cite{B10} (hereafter Paper I) offer models that are more thoroughly dynamical.
These models adopt a plausible model of the Galaxy's gravitational
potential $\Phi(R,z)$, in which there are substantial contributions to the local
acceleration from a disc, a bulge and a dark halo, all assumed to be
axisymmetric. They assume that motion parallel to the plane is
to some degree decoupled from motion perpendicular to the plane.
Specifically,  the vertical motion is governed by the time-dependent
potential
 \[\label{eq:slow}
\Psi_z(z;t)=\Phi[R(t),z]
\]
 where $R(t)$ is the radius at time $t$ that one obtains by assuming that the
radial motion is governed by the one-dimensional effective potential
 \[\label{eq:defsPsiR}
\Psi_R(R)=\Phi(R,0)+{L_z^2\over R^2}.
\]
 Paper I assumed that the
time-dependence of the potential (\ref{eq:slow}) is slow enough
that the action $J_z$ of vertical motion is constant. It justified this
assumption by referring to Figure 3.34 in \cite{BT08} (hereafter BT08), which
shows that the boundaries of one particular orbit are fairly well recovered
by the adiabatic approximation (hereafter \aa).  In this paper we explore the
validity of the \aa\ much more extensively. Our other goal is to present a
model of the Galactic disc that is not reliant on the \aa. This model
dispenses with the assumption that the $R$ and $z$ motions are decoupled by
using numerically synthesised orbital tori. 

The paper is organised as follows. In Section \ref{sec:test} the validity of
the \aa\ is tested on typical orbits. In Section \ref{sec:tori} we explain
the general principles of torus modelling and why we believe this technique
will prove a valuable tool for the interpretation of observational data.  We
define the torus-based model of the Galactic discs that we will use to test
the accuracy of observables obtained from the \aa, and we summarise the
methods used to extract observables when the model is based on (i) tori, and
(ii) the adiabatic approximation.  In Section \ref{sec:comp} we compare the
model's observables with estimates of them obtained from the \aa. In Section
\ref{sec:tilt} we examine the tilt of the velocity ellipsoid near the Sun,
which cannot be computed from the \aa, and show that its long axis points
towards the Galactic centre even though the potential is significantly
flattened. Section \ref{sec:conclude} sums up and looks to the future. An
appendix explains how some important Jacobians are calculated for a torus
model.

\begin{figure}
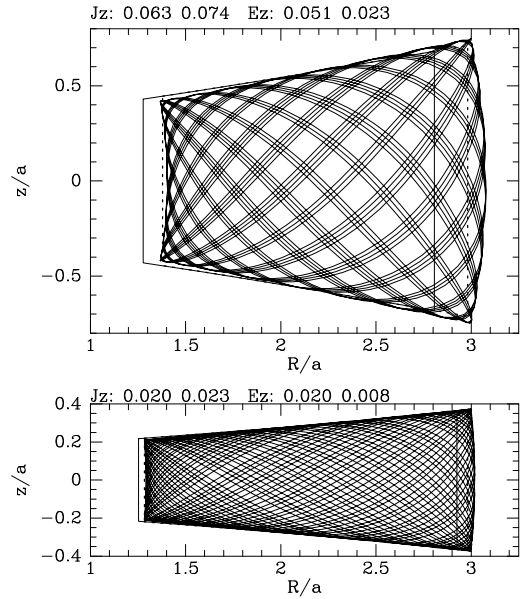

\centerline{\epsfig{file=graph1.5/miyam0.ps,width=.8\hsize}}
\vskip5pt
\centerline{\epsfig{file=graph1.5/miyam1.ps,width=.8\hsize}}
 \caption{Two orbits in the meridional plane of a Miyamoto-Nagai model
with scale-length ratio $b/a=0.2$. In both cases the angular momentum about the
symmetry axis is $L_z=\sqrt{GM/a}$, where $M$ is the model's mass. In  units of
$\sqrt{GMa}$, the actions of the upper and lower orbits are
$(J_r,J_z)=(0.109,0.067)$ and $(0.127,0.022)$.
The numbers above each panel are the values of $J_z$ of obtained by following the
motion of particles dropped from the upper left and upper right corners of
the orbit in the one-dimensional potential $\Psi_z(z)=\Phi(R,z)-\Phi(R,0)$,
and the corresponding vertical energies.  The nearly straight full lines show the \aa\ to the
orbit when $J_z$ is set to the average of the values at top left and the
radial action takes its true value. The dashed lines show the  boundary
yielded by the \aa\  when $L_z$ is replaced in the effective potential by
$L_z+J_z$.}\label{fig:expdisc}
\end{figure}

\begin{figure*}
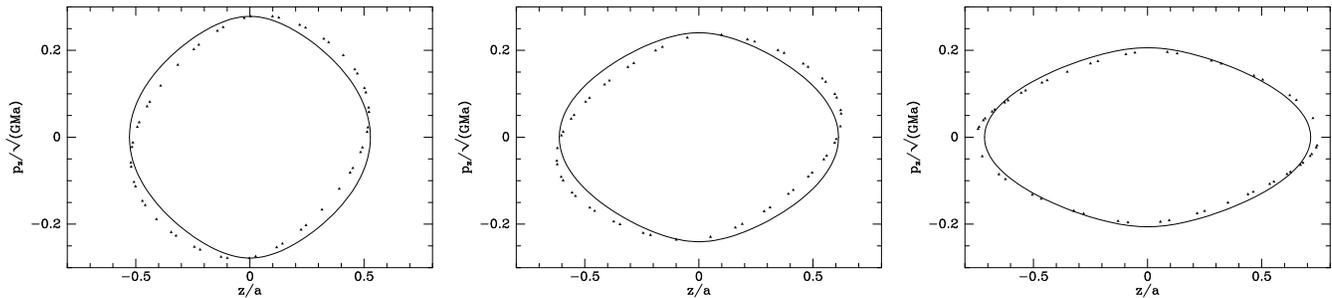

\centerline{\epsfig{file=graph1.5/miyamsos1.ps,width=.32\hsize}\quad
\epsfig{file=graph1.5/miyamsos2.ps,width=.32\hsize}\quad
\epsfig{file=graph1.5/miyamsos3.ps,width=.32\hsize}}
\caption{Surfaces of section $R/a=1.9,\,2.4$ and $3$ for the orbit shown in
the top panel of \figref{fig:expdisc}. The points are for the numerically
integrated orbit, while the curves are obtained from  the \aa.}\label{fig:expsos}
\end{figure*}

\section{validity of the adiabatic approximation}\label{sec:test}

Each panel of \figref{fig:expdisc} shows an orbit in the meridional plane in
the gravitational potential of a Miyamoto-Nagai Galaxy model
\citep{MiyamotoN} with scale-length ratio $b/a=0.2$: Figure 2.7 of BT08 shows
that this model  has a prominent
disc. From the Fourier transforms of the time series $R(t)$
$z(t)$ on these orbits \citep{BinneyS} we find that in units of
$\sqrt{GMa}$ their actions are $(J_r,J_z)=(0.109,0.067)$ and
$(0.127,0.022)$, respectively.  One can also estimate the vertical actions
$J_z$ of these orbits by dropping particles from points on their upper edges,
and following their motion in the one-dimensional potential (\ref{eq:slow})
with $R$ frozen at its current value -- see equation (\ref{eq:actints}) below.
The numbers at top of each panel show the
values obtained for $J_z$ in the same units when particles are dropped from
the top left and top right corners of the orbit; the values are displaced
from the true value by $\lta7\%$.  The corresponding vertical energies $E_z$
are also shown at the top of each panel; they differ by more than a factor 2.
Thus the \aa\ does provide a fairly good guide to how the vertical motion is
influenced by the radial motion.

The nearly straight full lines in \figref{fig:expdisc} show the outlines of
the approximate orbits that we obtain from the \aa\ by setting $J_r$ to its true
value and $J_z$ to the average of the values given at the top of the panel.
The shape of each approximate orbit is reasonable, although the left and
right edges are straight rather than curved, but the orbit is clearly
displaced to smaller radii with respect to the true orbit.  This difference
reflects the fact that the vertical motion contributes to the centrifugal
potential alongside the azimuthal motion. By assuming that the radial motion
is governed by the effective potential (\ref{eq:defsPsiR}) in which $L_z$
occurs rather than the total angular momentum $L$, we have under-estimated
the centrifugal potential.  Consequently, we predict that the orbit lives at
smaller radii than it really does.

In a spherical potential, the total angular momentum is related to $L_z$
and $J_z$ by $L=|L_z|+J_z$ (e.g.\ BT08 \S3.5.2)
and the radial motion is governed by an effective
potential in which the centrifugal component is $L^2/2r^2$, where
$r^2=R^2+z^2$. Consequently, an
obvious strategy for improving the predictions of the \aa\ is to replace $L_z$
in the effective potential by $L+J_z$.  The
dashed lines in \figref{fig:expdisc} show the effect of replacing $L_z$ by
 \[\label{eq:cfac}
\cL_z\equiv |L_z|+\gamma J_z,
\]
 with $\gamma=1$. Both orbits are now quite closely modelled. 

If we calculate the radial action of a given phase-space point $(\vx,\vv)$
using  $L_z$ rather than $\cL_z$ in the effective potential, the value we
obtain is too large when $(\vx,\vv)$ lies near apocentre, because the star
moves in an effective potential that has its minimum at a radius that
is too small. Conversely, when $(\vx,\vv)$ lies near pericentre, our estimate
of $J_r$ is too small if we use $L_z$. Since the \df\ decreases with
increasing $J_r$, the use of the less accurate
effective potential leads to phase-space points near pericentre being
over-weighted relative to points near apocentre, and this in consequence
shifts the predicted distribution in $v_\phi$ to large values. Hence,
replacing $L_z$ in the effective potential with $\cL_z$ for suitably chosen
$\gamma$ can usefully improve the accuracy of results obtained with the \aa.

The points in \figref{fig:expsos} show the consequents of the upper orbit of
\figref{fig:expdisc} in three surfaces of section that are obtained by noting
$z$ and $p_z$ when the star crosses the line $R=\hbox{constant}$ in the
meridional plane with $p_R>0$. The curves in each panel show the dependence
of $p_z$ on $z$ along the one-dimensional orbit in the potential $\Phi(R,z)$
with $R$ fixed at the appropriate value when the action $J_z$ is set to the
average of the values given above the top panel of \figref{fig:expdisc}. The
agreement between the curves yielded by the \aa\ and the numerical
consequents is on the whole good. In the left and
central panels we see that while the curves have reflection symmetry in
$p_z=0$ the consequents do not. This is because the surface of section is for
$p_R>0$, and when the star is moving outwards, it is likely to be
moving upwards when $z>0$ and downwards when $z<0$. As we will discuss
in Section \ref{sec:tilt}, when a galaxy is
formed out of such orbits, this $z$-dependent correlation between $p_R$ and
$p_z$ causes the principal axes of the velocity ellipsoid to become inclined
to the $R,z$ axes at $|z|>0$.  The \aa\ is unable to capture this aspect of the
dynamics and will always yield a velocity ellipsoid that is aligned with the
$R,z$ axes.

The panel on the extreme right shows that at large radii the \aa\
underestimates the maximum height $z_{\rm max}$ reached by a star, although
at most values of $z$ it predicts $p_z$ with good accuracy. 

The analogue of \figref{fig:expsos} for the orbit shown in the lower panel of
\figref{fig:expdisc} shows smaller offsets between the numerical
consequents and the predictions of the \aa, because the latter works best for
small vertical amplitudes.

With $v_{\rm t}$ denoting the tangential speed, the centrifugal potential is
$v_{\rm t}^2/r^2$. In a separable potential, the time average of the $i$th
component of velocity is related to the $i$th frequency and action by
$\langle v_i^2\rangle=\Omega_iJ_i$, so when we replace $v_{\rm t}^2$ by its
time average the centrifugal potential becomes
 \[
\sim\Omega_\phi{ |L_z|+(\Omega_z/\Omega_\phi)J_z\over R^2+z^2}.
\]
 The standard \aa\ underestimates the centrifugal potential by neglecting the
term proportional to $J_z$ in the numerator, and partially compensates by
neglecting the $z^2$ in the denominator. This neglect of $z^2$ must be
responsible for the fact that we find the optimum value of $\gamma$ to be
unity rather than $2$, which is a typical value of $\Omega_z/\Omega_\phi$ for
disc stars at $R_0$ in plausible Galactic potentials. However, in an
unrealistically flat potential, larger values of $\gamma$ prove optimal. For
example, when the potential is that of a razor-thin exponential disc and
there is no contribution from a bulge or a dark halo, we find
$\gamma\simeq1.9$ is required. Even in this case $\gamma$ is smaller than the
typical value of $\Omega_z/\Omega_\phi$ on account of the neglect of $z^2$ in
the denominator.

\section{A model based on orbital tori}\label{sec:tori}

The classical approach to modelling globular clusters starts by positing an
analytic form for the distribution function (\df) and then calculating the
density distribution and kinematics that are implied by this \df. Thus
globular clusters have been successfully modelled with \df s of the
King--Michie form (e.g.\ BT08 \S4.3). This approach can be extended to disc
galaxies.  For example \cite{Rowley88} modelled S0 galaxies with distribution
functions of the form $f(E,L_z)$, where $E$ and $L_z$ are, respectively,
orbital energy per unit mass and and angular momentum per unit mass about the
symmetry axis. Unfortunately, such a simple distribution function cannot
successfully model the Galaxy, because it predicts equal velocity dispersions
$\sigma_R$ and $\sigma_z$ in the radial and vertical directions, while
observations show that $\sigma_R\simeq1.7\sigma_z$ \citep[e.g.][]{AumerB09}.
A successful \df\ for the Galaxy must depend, explicitly or otherwise, on the
vertical action $J_z$.

Given that the \df\ will depend on two of an orbit's three actions $J_r$,
$J_z$ and $L_z$, substantial advantages arise from employing $J_r$ as the
other argument of the \df\ in place of $E$. For this reason Paper I
studied Galaxy models in which each stellar component had a \df\ that was an
analytic function of the three actions. It used the \aa\
to calculate observables from the \df. The purpose of this section is to
compare observables obtained in this way to those  obtained without
invoking the \aa\ but instead using orbital tori.

\subsection{General principles of torus modelling}

Orbital tori are the three-dimensional surfaces in six-dimensional phase
space on which individual orbits move. They are the building blocks from
which Jeans' theorem assures us that any equilibrium model can be built. Each
torus is characterised by a set of three actions $\vJ=(J_r,L_z,J_z)$ and
therefore corresponds to a point in action space. We build a galaxy model by
assigning a weight to each torus. 

We obtain tori as the images of analytic tori under a canonical
transformation. The tori used here are defined by the angle-action
coordinates of the isochrone potential (e.g.\ BT08 \S3.5). Given actions
$\vJ$, the computer constructs a canonical transformation that maps the
analytic torus with actions $\vJ$ into the required torus by adjusting the
coefficients in a trial generating function so as to minimise the rms
variation of the Galactic Hamiltonian on the image torus. Once this has been
done, we have analytic expressions for the phase-space coordinates
[$\vx(\vtheta),\vv(\vtheta)$] as functions of the angle variables $\theta_i$,
which control the orbital phase. On a given torus, the phase-space
coordinates $(\vx,\vv)$ are $2\pi$-periodic functions of each $\theta_i$.
The torus-fitting program also returns the values of the torus's
characteristic frequencies $\Omega_i$, so we can determine the motion of a
star using $\vtheta(t)=\vtheta(0)+{\bf\Omega} t$.  For a fuller summary of
how orbital tori are constructed and references to the papers in which torus
dynamics was developed, see \cite{McMB08}. 


Torus modelling is best understood as an extension of Schwarzschild modelling
\citep{SchwarzI}, which has been successfully used in many studies of the
dynamics of external galaxies \citep[e.g.][]{Gebhardt03,Cappellari}.  A
Schwarzschild model is constructed by assigning weights to each orbit in a
``library'' of orbits. The orbit library is assembled by integrating the
equations of motion in the given potential for a sufficient time, and noting
the fraction of its time that the orbiting particle spends in each bin in the
space of observables. Then a non-negative weight $w_i$ is chosen for each
orbit such that the data are consistent with the model's predictions. 
In torus modelling orbits are replaced by tori, which are essentially
equivalence classes of orbits that differ from one another only in phase, and
a Runge-Kutta integrator is replaced by the torus-fitting code.  Whereas
orbits are defined by their six-dimensional initial conditions, tori are
defined by their actions $\vJ$.

Replacing numerically integrated orbits with orbital tori brings the following
advantages

\begin{enumerate}

\item{} The phase-space density of orbits becomes known because tori have
prescribed actions and the six-dimensional phase-space volume occupied by
orbits with actions in $\d^3\vJ$ is $\tau=(2\pi)^3\d^3\vJ$. Knowledge of the
phase-space density of orbits allows one to convert between  orbital
weights and the value taken by the \df\ on an orbit.

\item{} On account of the above result, there is a clean and unambiguous
procedure for sampling orbit space and relating the weights of individual
tori to the value that the \df\ takes on them -- see below. The choice of
initial conditions from which to integrate orbits for a library is less
straightforward because the same orbit can be started from many initial
conditions, and when the initial conditions are systematically advanced
through six-dimensional phase space, the resulting orbits are likely at some
point to cease exploring a new region of orbit space and start resampling a
part of orbit space that is already represented in the library. On account of
this effect, it is hard to relate the weight of an orbit to the value taken
by the \df\ on it \citep[but
see][]{Haefner,Thomas05}.

\item{} There is a simple relationship between the distribution of stars in
action space and the observable structure and kinematics of the model; as
explained in \S4.6 of BT08, the observable properties of a model
change in a readily understood way when stars are moved within action space.
The simple relationship between the observables and the distribution of stars
in action space enables us to infer from the observables the form of the
\df\ $f(\vJ)$, which is nothing but the density of stars in action space.

\item{} From a torus one can readily find the velocities that a star on a
given orbit can have when it reaches a given spatial point $\vx$. By contrast
a finite time series of an orbit is unlikely to exactly reach $\vx$, and
searching for the time at which the series comes closest to $\vx$ is
laborious. Moreover, several velocities are usually possible at a given
location, and a representative point of closest approach must be found for
each possible velocity.

\item{}
An orbital torus is represented by of order 100 numbers while a
numerically-integrated orbit is represented either by some thousands of
six-dimensional phase-space locations, or by a similar number of occupation
probabilities within a phase-space grid.

\item{} The numbers that characterise a torus are smooth functions of the
actions $\vJ$. Consequently tori for actions that lie between the points of
any action-space grid can be constructed by interpolation on the grid.
Interpolation between time series is not practicable.

\item{} Schwarzschild and torus models are zeroth-order, time-independent
models which differ from real galaxies by suppressing time-dependent
structure, such as ripples around early-type galaxies
\citep{MalinC,QuinnP,SchweizerF}, and spiral structure or warps in disc
galaxies. Since the starting point for perturbation theory is action-angle
variables \cite[e.g.][]{Kalnajs}, in the case of a torus model one is well
placed to add time-dependent structure as a perturbation.
\cite{Kaasalainen95a} showed that classical perturbation theory works
extremely well when applied to torus models because the integrable
Hamiltonian that one is perturbing is typically much closer to the true
Hamiltonian than in classical applications of perturbation theory
\citep{GerhardS,DehnenG,Weinberg}, in which the unperturbed Hamiltonian
arises from a potential that is separable (it is generally either spherical
or plane-parallel).

\end{enumerate}

\subsection{Choice of the DF}

For the comparison of results obtained with and without the \aa\ it is
appropriate to study a model that has a very simple \df.  Specifically we
represent both the thin and the thick discs with a \df\ that is
quasi-isothermal in the sense of Paper I:
\[\label{totalDF}
f(J_r,L_z,J_z)=f_{\sigma_r}(J_r,L_z)\times
{\nu_z
\over2\pi\sigma_z^2}\,\e^{-\nu_z J_z/\sigma_z^2},
\]
where
 \[\label{planeDF}
f_{\sigma_r}(J_r,L_z)\equiv{\Omega\Sigma\over\pi\sigma_r^2\kappa}\bigg|_{\Rc}
[1+\tanh(L_z/L_0)]\e^{-\kappa J_r/\sigma_r^2}.
\]
 Here $\Omega(L_z)$ is the circular frequency for angular momentum $L_z$,
$\kappa(L_z)$ is the radial epicycle frequency and $\nu(L_z)$ is its vertical
counterpart.  $\Sigma(L_z)=\Sigma_0\e^{-(R-\Rc)/R_\d}$ is the (approximate)
radial surface-density profile, where $\Rc(L_z)$ is the radius of the
circular orbit with angular momentum $L_z$. The factor $1+\tanh(L_z/L_0)$ in
equation (\ref{planeDF}) is there to effectively eliminate stars on
counter-rotating orbits and the value of $L_0$ is unimportant provided it is
small compared to the angular momentum of the Sun. In equations
(\ref{totalDF}) and (\ref{planeDF}) the functions $\sigma_z(L_z)$ and
$\sigma_r(L_z)$ control the vertical and radial velocity dispersions. The
observed insensitivity to radius of the scaleheights of extragalactic discs
motivates the choices
 \begin{eqnarray}\label{eq:sigmas}
\sigma_r(L_z)&=&\sigma_{r0}\,\e^{q(R_0-\Rc)/R_\d}\nonumber\\
\sigma_z(L_z)&=&\sigma_{z0}\,\e^{q(R_0-\Rc)/R_\d},
\end{eqnarray}
 where $q=0.45$ and $\sigma_{r0}$ and $\sigma_{z0}$ are approximately equal
to the radial and vertical velocity dispersions at the Sun. We take the \df\
of the entire disc to be the sum of a \df\ of the form (\ref{totalDF}) for
the thin disc, and a similar \df\ for the thick disc, the normalisations
being chosen so that at the Sun the surface density of thick-disc stars is
23 per cent of the total stellar surface density. Table \ref{tab:df} lists the
parameters of each component of the \df.

\begin{table}
\caption{Parameters of the \df.}\label{tab:df}
\begin{center}
\begin{tabular}{l|cccc}
Disc & $R_\d/\hbox{kpc}$ & $\sigma_{r0}/\!\kms$ & $\sigma_{z0}/\!\kms$ &
$L_0/\!\kpc\kms$\\
\hline
Thin & 2.4 & 27 & 20 & 10\\
Thick& 2.5 & 48 & 44 & 10\\
\end{tabular}
\end{center}
\end{table}

There are two main differences between the \df\ we use here and that used in
Paper I: (i) Paper I used the actual vertical frequency $\Omega_z(\vJ)$ in
equation (\ref{totalDF}) while here we use the vertical epicycle frequency
$\nu(L_z)$.  This substitution is necessary because for large $\vJ$,
$\Omega_z$ tends to zero so fast that the product $\Omega_z J_z$ can decrease
as $J_z\to\infty$, leading to unphysical results when $\Omega_zJ_z$ appears
in the \df\ as the argument of an exponential. (ii) In the interests of
simplicity the thin disc is here represented by a single quasi-isothermal
component whereas in Paper I it was represented it by a sum of
quasi-isothermals, one for stars of each age. 

Any serious attempt to fit a real stellar catalogue must distinguish between
stars of different ages, and different metallicities, because the colours and
luminosities of stars are very much functions of age and metallicity, so the
chances of a star entering a catalogue depend on its age and metallicity.
Consequently, by lumping together all thin-disc stars regardless of age we
forgo the opportunity to fit a real stellar catalogue in a detailed way.
Nonetheless, we shall require our \df\ to reproduce an observational density
profile to demonstrate that even our unrealistically simple \df\ has
sufficient flexibility to reproduce given data to reasonable precision.  

The physical properties of the model are jointly determined by the \df\ and
the gravitational potential $\Phi(R,z)$. Ultimately it will be necessary to
require that the Galaxy's \df\ be consistent with $\Phi$ in the sense that
the density of matter that the \df\ predicts generates $\Phi$. However,
before the question of dynamical self-consistency can be addressed, one must
not only specify the \df\ of dark matter (which is believed to contribute
about half the gravitational force on the Sun) but also distinguish carefully
between the masses of stars and their luminosities in the wavebands in which
they are observed. In practice the latter can be done only if one has
specified the Galaxy's star-formation and metal-enrichment history. This
enterprise goes far beyond the scope of the present paper; it will be
addressed in subsequent papers in this series, which will explain the
importance of comparing models to data in the space of observables, such as
apparent magnitudes, parallaxes and proper motions, rather than the space of
physical variables such as $(\vx,\vv)$ used here. The purpose of this paper
is merely to lay the foundations for such an exercise, which we expect to
give first insights into the \df\ of dark matter.  Here we take the view that
our \df\ weights stars by their luminosity rather than their mass, and assume
that $\Phi$ is the potential of Model 2 of \cite{DehnenB97} modified to have
thin- and thick-disc scaleheights of $360\pc$ and $1\kpc$
(Table~\ref{tab:pot}). In this model the disc contributes 60 per cent of the
gravitational force on the Sun, with dark matter contributing most of the
remaining force.

\subsection{Modelling procedures}

Together the \df\ and the potential specify the probability density of stars
in phase space. The simplest way to derive the model's physical
characteristics from this probability density is to obtain a discrete
realisation of the probability density by Monte-Carlo sampling. The model's
physical characteristics are then obtained by binning the realisation's
stars. The \df\ specified by equations (\ref{totalDF}) and (\ref{planeDF})
can be analytically integrated over $J_z$ and $J_r$ to obtain the marginal
distribution in $L_z$, so we can obtain a discrete realisation of this \df\
by successively sampling one-dimensional pdfs in $L_z$, $J_r$ and $J_z$. The
results presented below are typically obtained with $\sim200\,000$ tori.

Once we have a torus library, a discrete realisation of the Galaxy is
obtained by repeatedly choosing a torus from the library at random, then
choosing each angle variable uniformly within $(0,2\pi)$, and using the
functions returned by the torus-generating software to to determine
$(\vx,\vv)$ from the given values of $\vJ$ and $\vtheta$.

When the \aa\ is used, model construction proceeds rather
differently: then given $(\vx,\vv)$ one determines $J_r$ and $J_z$ by the
following steps.

\begin{table}
\caption{Parameters of the potential}\label{tab:pot}
\begin{center}
\begin{tabular}{l|cccccc}
\hline
Component & $\Sigma(R_0)/\msun \pc^{-2}$ & $R_\d/\hbox{kpc}$ & $h/\!\kpc$ & $R_{\rm m}/\!\kpc$\\
\hline
Thin &36.42& 2.4 & 0.36 & 0\\
Thick&4.05& 2.4 & 1 & 0\\
Gas &8.36& 4.8 & 0.04 & 4\\
\hline
\end{tabular}
\begin{tabular}{l|cccccc}
Component & $\rho/\!\msun\pc^{-3}$ & $q$ & $\gamma$ & $\beta$ & $r_0/\!\kpc$
& $r_t/\!\kpc$\\
\hline
Bulge&0.7561&0.6&1.8&1.8&1&1.9\\
Halo&1.263&  0.8&$-2$&2.207&1.09&1000\\
\end{tabular}
\end{center}
\end{table}

\begin{figure}
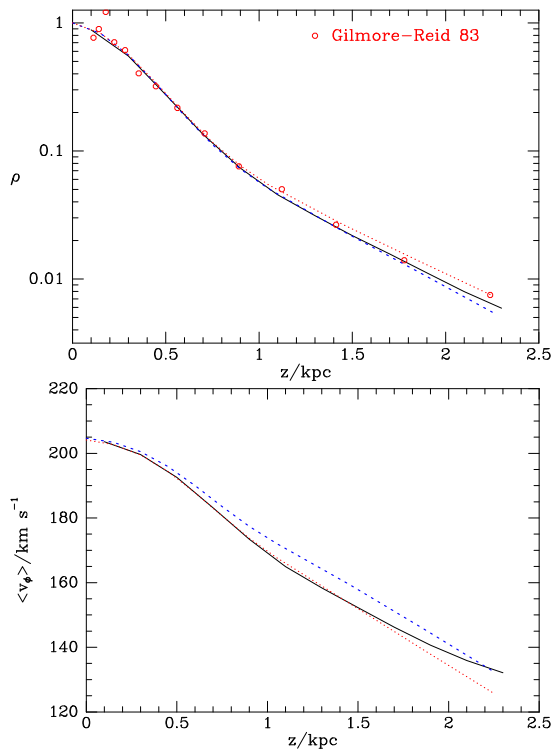

\begin{center}
\epsfig{figure=graph1.5/zrho_1.ps,width=.85\hsize}
\epsfig{figure=graph1.5/zvphi_1.ps,width=.85\hsize}
\end{center}
 \caption{Top: density as a function of distance from the midplane at the
solar radius. Bottom: mean streaming velocity as a function of distance from
the midplane.  The full black curves show the predictions of the full torus
model; the blue  curves are obtained from the \aa\ with $\gamma=0$; the dotted
red curves show the effect of setting $\gamma=1$.}\label{fig:rhoz}
\end{figure}

\begin{figure}
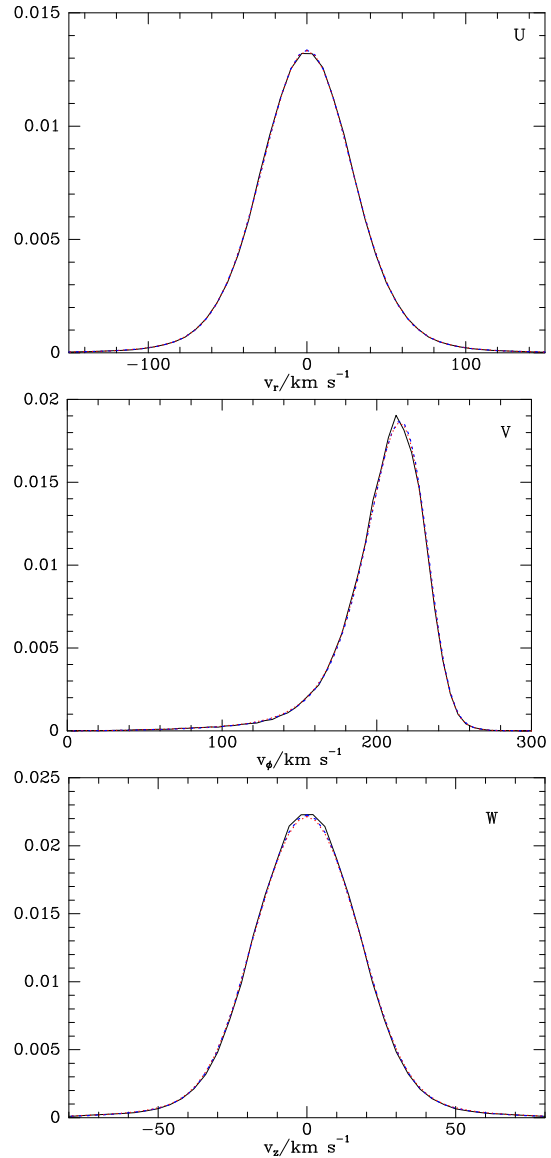

\begin{center}
\epsfig{figure=graph1.5/NU_0.ps,width=.85\hsize}
\epsfig{figure=graph1.5/NV_0.ps,width=.85\hsize}
\epsfig{figure=graph1.5/NW_0.ps,width=.85\hsize}
\end{center}
 \caption{The distributions for local stars of the radial, azimuthal and
vertical components of velocity after marginalising over the other two
components. The full black curves are for the torus-based
model while the broken curves are obtained using
the \aa\ with $\gamma=0$ (dashed blue) and $\gamma=1$ (dotted red).
The curves overlie one another too closely to be clearly distinguishable.}\label{fig:UVW}
\end{figure}

\begin{itemize}
\item Evaluate the vertical and radial energies
 \begin{eqnarray}
E_z&=&\fracj12v_z^2+\Psi_z(z)\nonumber\\
E_R&=&\fracj12 v_R^2+\Psi_R(R),
\end{eqnarray}
 where $\Psi_z$ and $\Psi_R$ are defined by equations (\ref{eq:slow}) and
(\ref{eq:defsPsiR}).

\item Evaluate the actions from 
\begin{eqnarray}\label{eq:actints}
J_z={2\over\pi}\int_0^{z_{\rm max}}\d z\,v_z(E_z,z)\nonumber\\
J_r={1\over\pi}\int_{R_{\rm p}}^{R_{\rm a}}\d R\,v_R(E_R,R),
\end{eqnarray}
 where $z_{\rm max}$ is defined by $\Psi_z(z_{\rm max})=E_z$ and $R_{\rm a}$
and $R_{\rm p}$ are defined by $\Psi_R(R_i)=E_R$.

\end{itemize}
 These steps make it straightforward to evaluate the \df\ at an arbitrary
point $(\vx,\vv)$, and thus derive the model's physical properties by
numerically integrating the \df, times any power of $v_i$, over velocity
space. The quantities such as stellar number density $\nu(\vx)$ and velocity
dispersion $\langle v_z\rangle^{1/2}(\vx)$ are then continuous functions of their arguments.
In the absence of the \aa, an iterative procedure such as that described by
\cite{McMB08} is required to determine $\vJ(\vx,\vv)$, and the
torus-modeeling procedure avoids this procedure by choosing $\vJ$ not
$(\vx,\vv)$. The price we pay for starting with $\vJ$ is discreteness, and
the necessity of estimating $\nu(\vx)$, etc, by binning stars.

\begin{figure}
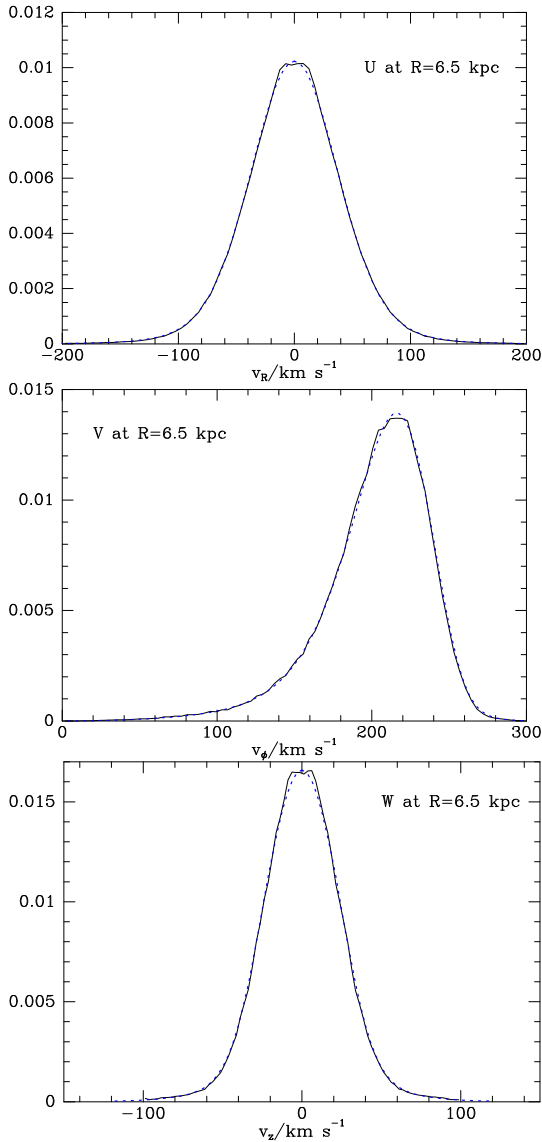

\centerline{\epsfig{file=graph1.5/NU_R6.5.ps,width=.85\hsize}}
\centerline{\epsfig{file=graph1.5/NV_R6.5.ps,width=.85\hsize}}
\centerline{\epsfig{file=graph1.5/NW_R6.5.ps,width=.85\hsize}}
\caption{As \figref{fig:UVW} except for a volume that lies in the plane at
$R=6.5\kpc$.}
\label{fig:UVW6.5}
\end{figure}

\begin{figure}
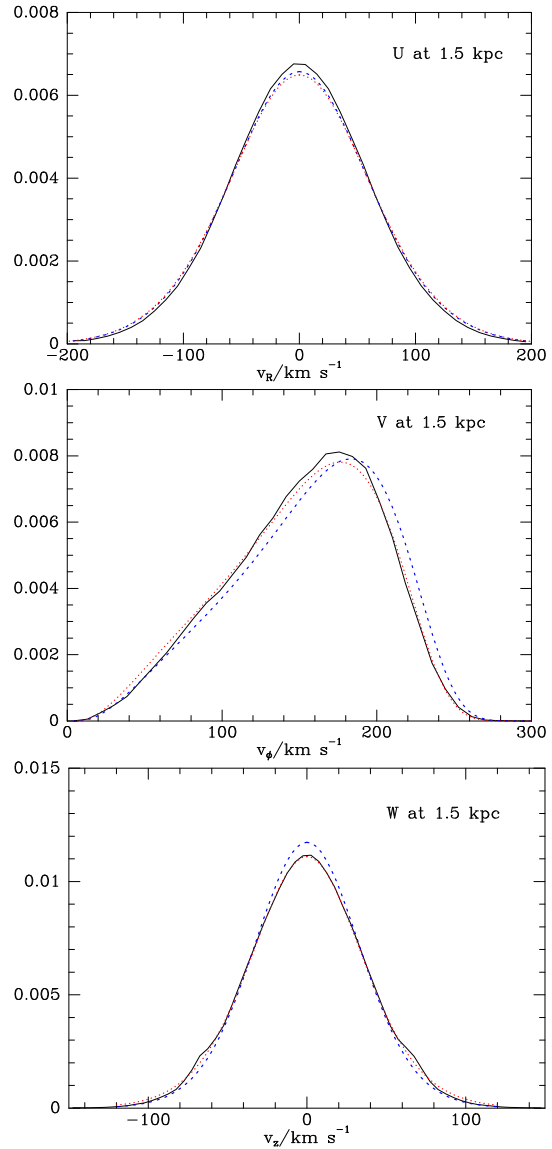

\centerline{\epsfig{file=graph1.5/NU_1.5.ps,width=.85\hsize}}
\centerline{\epsfig{file=graph1.5/NV_1.5.ps,width=.85\hsize}}
\centerline{\epsfig{file=graph1.5/NW_1.5.ps,width=.85\hsize}}
\caption{As \figref{fig:UVW} except for a volume that is $1.5\kpc$ from the
plane. The dotted red curves are for the \aa\ with
$\gamma=1$ rather than 0, shown by the broken blue curves.}
\label{fig:UVWhigh}
\end{figure}

\begin{figure}
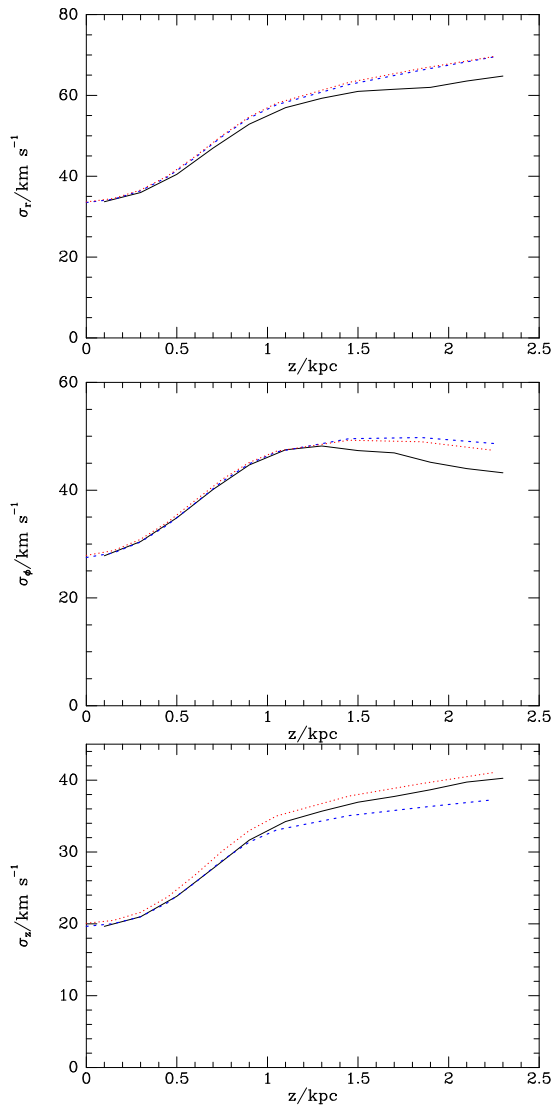

\begin{center}
\epsfig{figure=graph1.5/zsr_1.ps,width=.85\hsize}
\epsfig{figure=graph1.5/zsphi_1.ps,width=.85\hsize}
\epsfig{figure=graph1.5/zsz_1.ps,width=.85\hsize}
\end{center}
\caption{The radial, azimuthal and vertical velocity dispersions as
functions of distance from the midplane. The full black curve shows results
from the torus model, while the dashed blue line is obtained using the \aa\
with $\gamma=0$; the dotted red curves show the effect of setting $\gamma=1$.}
\label{fig:sigs}
\end{figure}

\section{Comparisons}\label{sec:comp}

\figref{fig:rhoz} shows the density of stars (upper panel) and
the mean-streaming velocity (lower panel) as functions of $|z|$ at the solar
radius. The  full curves show results from the torus model, while the dashed
and dotted curves show results obtained from the \aa\ using two values of the
parameter $\gamma$ defined by eq.~(\ref{eq:cfac}): $\gamma=0$ (dotted red) and
$\gamma=1$ (dashed blue). Also shown in the upper panel are the seminal data
points of \cite{GilmoreR}, which led to the identification of the thick
disc. Since our \df\ provides a reasonable fit to these points, it may be
close to the actual \df\ for turnoff stars. The \aa\ recovers the density
profile of the torus model to good accuracy for either value of $\gamma$.

The lower panel in \figref{fig:rhoz} shows how the mean streaming speed
$\ex{v_\phi}$ is predicted to fall with $|z|$. The agreement between the torus
model and the model based on the \aa\ with $\gamma=1$ (dotted red) is excellent
for $|z|\la1.6\kpc$ but at larger heights the \aa\ has a systematic tendency to
under-estimate $\ex{v_\phi}$. On account of the problem discussed in 
Section \ref{sec:test} apropos \figref{fig:expdisc}, the \aa\ with $\gamma=0$ over-estimates
the mean-streaming speed by a few $\!\kms$ for $|z|\gta0.7\kpc$.

\figref{fig:UVW} shows the distributions of the radial ($U$), tangential
($V$) and vertical ($W$) components of velocity in a small volume akin to the
solar neighbourhood. Since the full black curves from the torus model
coincide with the dashed blue curves from the \aa\ with $\gamma=0$, the \aa\
reproduces the torus results to high precision. The results obtained on
setting $\gamma=1$ are plotted as a dotted red curve but overlie the other
curves too closely to be clearly distinguishable.
\figref{fig:UVW6.5} shows that the \aa\ is equally successful in recovering the
distributions of $U$, $V$ and $W$ at a smaller Galactocentric radius,
$6.5\kpc$. The insensitivity to $\gamma$ of the velocity distribution in the
plane arises because these distributions are dominated by rather nearly
circular orbits, and for these orbits $J_z$ is so much smaller than $L_z$
that adding $J_z$ to $L_z$ barely changes the numerical value.

\figref{fig:UVWhigh} shows the $U$, $V$ and $W$ distributions at $R=8\kpc$
and $1.5\kpc$ away from the midplane. Systematic differences between the
predictions of the \aa\ and the results of the full torus model are now
evident. For either value of $\gamma$, the \aa\ yields a distribution of $U$
that is slightly too broad. When $\gamma=0$, the \aa\ gives a distribution in
$W$ (dashed blue curve) that is too sharply peaked, but this fault is nicely
corrected by setting $\gamma=1$ (dotted red curve) because increasing
$\gamma$ moves orbits of given $L_z$ outwards, and by virtue of equation
(\ref{eq:sigmas}), the smaller an orbit's value of $L_z$, the faster it is
likely to move vertically.  As expected, with $\gamma=0$ the \aa\ yields a
distribution in $V$ that is offset from that of the full torus model by
$\sim8\kms$ towards higher velocities (dashed blue curve). This offset is
largely cured by setting $\gamma=1$ (dotted red curve).

\figref{fig:sigs} shows the variation with $|z|$ of the radial, tangential
and vertical velocity dispersions. The two planar velocity dispersions are
accurately reproduced for $|z|\la1.3\kpc$. At greater heights the \aa\
over-estimates the dispersions, most strikingly so in the case of
$\sigma_\phi$.  The excessive value of $\sigma_\phi$ is clearly associated
with the tendency of the red curve in the middle panel of
\figref{fig:UVWhigh} to lie above the black one at $v_\phi\sim50\kms$.  As
the top and middle panels of \figref{fig:UVWhigh} suggest, $\langle
v_R^2\rangle^{1/2}$ and $\langle (v_\phi-\langle v_\phi\rangle)^2\rangle^{1/2}$ are
at any height remarkably insensitive to the value of $\gamma$. 

From the bottom panel of \figref{fig:UVWhigh} we anticipate that increasing
$\gamma$ will increase $\langle v_z^2\rangle$ at large $|z|$ and indeed the
bottom panel of \figref{fig:sigs} shows that increasing $\gamma$ from zero to
unity increases $\sigma_z$ at all $z$, but particularly at large $z$.  This
result arises because increasing $\gamma$ shifts orbits with large $J_z$
outwards, and since $f(\vJ)$ is a strongly
decreasing function of $|\vJ|$, this outwards shift raises the density of
stars with large $J_z$ at a given location. At $|z|\ga 1\kpc$ setting
$\gamma=1$ increases the accuracy of $\sigma_z$, while at smaller values of
$|z|$, a slight deterioration in accuracy results. The bottom panel of
\figref{fig:UVWhigh} hints that the full curve in \figref{fig:sigs} may lie
slightly too low as a result of poor sampling by the torus model of orbits
with large $|v_z|$, so the results from the \aa\ with $\gamma=1$ may be more
accurate than appears to be the case.

\begin{figure}
\centerline{\epsfig{file=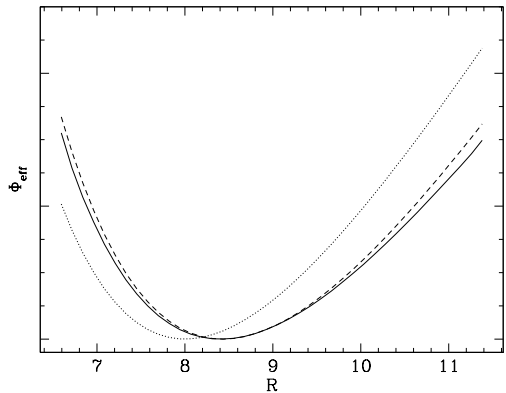,width=.65\hsize,angle=-90}}
\caption{Three effective potentials for the orbit in the Galaxy model that
has actions $(J_r,L_z,J_z)=(0.05,1,0.05)$ times the angular momentum of the
circular orbit at $R_0$. This orbit extends to $3\kpc$ above the plane. The
dotted and dashed curves show the naive effective potential
(\ref{eq:defsPsiR}) with $L_z$ replaced by $|L_z|+\gamma J_z$ and $\gamma=0$ in
the dotted case and $\gamma=1$ in the dashed case. The full curve shows the
effective potential derived from a time-average of the radial
force.}\label{fig:Phieff}
\end{figure}

The dotted and dashed curves in \figref{fig:Phieff} show the effective potential
(\ref{eq:defsPsiR}) for a typical solar-neighbourhood orbit when $\gamma=0$
and $\gamma=1$, respectively. The full curve shows an effective potential
for this orbit that was obtained by first evaluating the time-average of
$\p\Phi/\p R-L_z^2/R^3$ at each value of $R$ visited by the orbit, and then
integrating the resulting function of $R$. We see that with $L_z$ replaced by
$|L_z|+J_z$ the simple effective potential (\ref{eq:defsPsiR}) provides a good
fit to the effective potential obtained by time-averaging the radial force as
a function of $R$.

\begin{figure}
\centerline{\epsfig{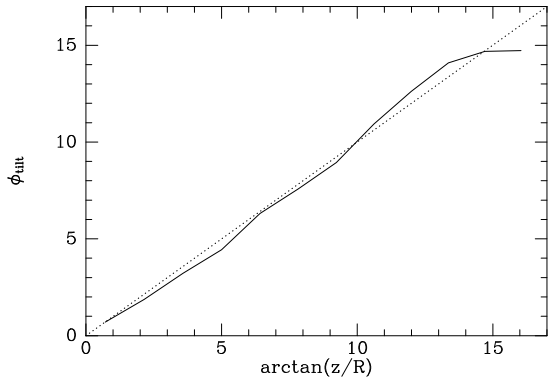}}
\caption{The variation of the angle of tilt of the velocity ellipsoid with
respect to the plane versus the angle $\arctan(|z|/R)$ between the plane and
the line of sight from ($R,z$) to the Galactic centre. Since the full curve tracks the
dotted line, the long velocity ellipsoid points almost straight at the
Galactic centre.}\label{fig:tilt_ang}
\end{figure}

\begin{figure}
\centerline{\epsfig{file=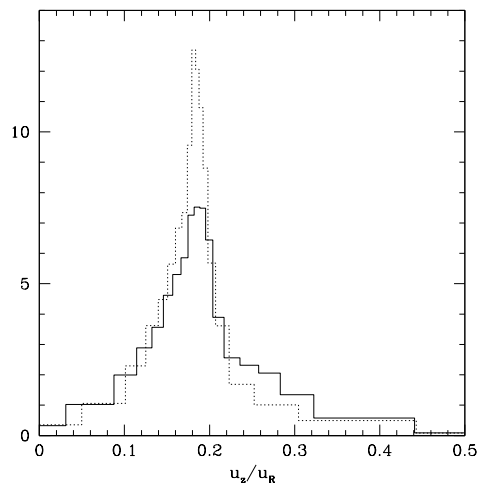,width=.8\hsize}}
\caption{The distribution of the ratio $u_z/u_R$ that gives the direction of
a principal axis of the contribution of each torus to the velocity ellipsoid
at the point $(R,z)=(R_0,1.5\kpc)$. The full histogram weights each torus
equally, while the dotted one weights them in proportion to their
contributions to the density at the given point. Even the latter distribution is
wide, so the orientation of the velocity ellipsoid depends quite sensitively
on the weights assigned to orbits by the \df.}\label{fig:tilt_hist}
\end{figure}

\section{Tilt of the velocity ellipsoid}\label{sec:tilt}

A popular diagnostic of the Galaxy's gravitational potential is the way in
which the principal axes of the velocity ellipsoid tilt as one moves away
from the plane \citep{Siebert08,Smith09}. As stated above, this
phenomenon lies beyond the scope of the \aa, but it can be determined from the
torus model. \figref{fig:tilt_ang} shows that this angle is nearly equal to
the angle $\arctan(|z|/R)$ between the plane and the line of sight from
($R,z$) to the
Galactic centre. That is, in the vicinity of the Sun, the longest axis of the
velocity ellipsoid is almost parallel to the radial vector $\vr$.

This behaviour is that expected in a spherical potential, and \cite{Smith09}
argue that alignment of the velocity ellipsoid with spherical coordinates
implies that the potential is spherically symmetric.  Our assumed
gravitational potential is far from spherical because roughly half the radial
force at the Sun is contributed by the disc, and the dark halo is itself
flattened (axis ratio $0.8$). The aspherical nature of the potential is
reflected in the fact that the distribution of frequency ratios
$\Omega_z/\Omega_\phi$ of the model's tori, has a median close to 2, while in
a spherical potential this ratio is inevitably unity. 

Although our model is not
strictly speaking a counter example to the assertion of \cite{Smith09}
because we have established only that the velocity ellipsoid is approximately
aligned with spherical coordinates in the region around the Sun, it does
suggest that one should examine more closely the reasoning in that paper.

A key step in its argument is the assertion that if the \df\ is an even
function of $v_r^2$, then the isolating integrals that constrain individual
orbits are also. If the isolating integrals have this property,
then the velocity ellipsoid provided by {\it any\/} \df\ will be aligned with
spherical coordinates. In particular the \df\ $f(\vJ)=\delta(\vJ-\vJ_0)$
corresponding to a single orbit will be radially aligned, so the matrix $\langle
v_iv_j\rangle_\vx$ will be diagonal in spherical coordinates, where
$\langle.\rangle_\vx$ implies the time average over instants when the star
lies in some small volume around $\vx$.

As \figref{fig:expdisc} illustrates, for stars on standard orbits in the
meridional plane, four velocities are possible at a given point. Let these
velocities be $\pm\vv_1$ and $\pm\vv_2$.  By time-reversal symmetry, the
probability that $\vv_1$ occurs is inevitably equal to the probability that
$-\vv_1$ occurs, and similarly for $\pm\vv_2$. However, it turns out that
$\pm\vv_1$ may occur more or less often than $\pm\vv_2$. In fact the
probability of occurrence of $\vv_i$ is proportional to the Jacobian
$\p(\vtheta)/\p(\vx)$, which is a non-trivial function of $\vtheta$, but one
that can be determined for a numerically constructed torus (see Appendix A).
The four possible velocities at $\vx$ correspond to the four values of
$\vtheta$ that bring the star to the given point $\vx$, and the values $p_1$
and $p_2$ taken by $\p(\vtheta)/\p(\vx)$ at these values of $\vtheta$ depend
on whether $\vv$ is $\pm\vv_1$ or $\pm\vv_2$.  Clearly we have
 \[
\langle v_iv_j\rangle_\vx={p_1v_{1i}v_{1j}+p_2v_{2i}v_{2j}\over p_1+p_2}.
\]
 Let $\vu$ be the eigenvector of this matrix that lies closest to the $\vr$
direction. \figref{fig:tilt_hist} shows for the point $(R,z)=(8,1.5)\kpc$ the
distribution of the ratio $u_z/u_R$. The full histogram shows the
distribution when each torus is given equal weight, while the dotted
histogram shows the distribution when tori are correctly weighted by the
contributions that they make to the stellar density at the given point.
(Although the density of sampling in action space ensures that all tori make
equal contributions to the stellar mass of the entire Galaxy, every torus has
its own way of spreading its mass in space.) We see that the distribution of
orientations of individual contributions to the velocity ellipsoid is quite
broad, even when the tori are correctly weighted. An examination of the
dependence of $u_z/u_R$ on $J_z$ reveals that orbits with larger values of
$J_z$ make contributions that are aligned with $\vr$, while it is orbits with
small $J_z$ that sometimes make contributions that are aligned nearly
parallel to the plane. Since the \df\ specifies the
relative weight of these variously oriented contributions, it controls the
orientation of the final velocity ellipsoid at least as much as does the
gravitational potential.  Consequently, only limited inferences about the
nature of the potential can be drawn from observations of the velocity
ellipsoid yielded by the \df\ that the Galaxy happens to have.

The widespread belief that the shape of the potential can be inferred from
the orientation of the velocity ellipsoid probably arises from studies of
models that have St\"ackel potentials. For these potentials the
Hamilton-Jacobi equation separates in an appropriate coordinate system
$(u,v)$, and the canonically conjugate momenta are functions of one
coordinate only: $p_u(u)$, $p_v(v)$ (e.g.\ BT08 \S3.5.3). Consequently, the
coordinate directions are bisectors of the angles between $\vv_1$ and
$\vv_2$. Moreover, it turns out that for these potentials,
$\p(\vtheta)/\p(\vx)$ is the same for all four values of $\vtheta$ so the
coordinate directions are the eigenvectors of $\langle v_iv_j\rangle_\vx$ for
every orbit that reaches $\vx$. Consequently, the velocity ellipsoid has to
be oriented with the coordinate directions regardless of how orbits are
weighted.\footnote{This conclusion can be obtained more simply by observing that the isolating integrals
$E$ and $I_3$ upon which the \df\ must depend, are even functions of $p_u$ and
$p_v$.} In a general potential, there is no universal coordinate system that
describes the alignment of the egenvectors of $\langle v_iv_j\rangle_\vx$,
and the orientation of the final velocity ellipsoid very much depends on how
orbits are weighted.

\section{conclusions}\label{sec:conclude}

Dynamical models of the Milky Way will play an key role in the scientific
exploitation of data from large surveys that are currently being undertaken.
Models that are based on Jeans' theorem should be the most powerful tools for
extracting science from data, and amongst such models those that express the
\df\ as a function of the actions enjoy some very important advantages. 

The major obstacle to the use of Jeans' theorem in the context of the Galaxy
is the lack of analytic expressions for three independent isolating
integrals. Paper I presented models that use approximate expressions for the
actions that rely on the adiabatic invariance of the vertical action $J_z$.
In Section \ref{sec:test} we tested the validity of this adiabatic
approximation (\aa) by numerically integrating typical orbits.  We found that
the orbits' vertical dynamics is reproduced by the \aa\ to remarkably good
accuracy, but the motion in the plane is less accurately recovered because
the naive \aa\ under-estimates the strength of the centrifugal potential.
This defect leads to the radial action derived for a given phase-space point
$(\vx,\vv)$ being over-estimated when the point lies near apocentre, and
under-estimated when it lies near pericentre. Since $v_\phi$ is small at
apocentre and large near pericentre and the \df\ is a declining function of
$J_r$, the defect leads to the mean-streaming velocity $\langle
v_\phi\rangle$ being over-estimated. The problem can be largely resolved by
replacing $L_z$ in the centrifugal potential by $|L_z|+\gamma J_z$ with
$\gamma$ a number of order unity.

The more strongly flattened the potential is, the more accurate the \aa\
becomes and the larger the value of $\gamma$ needs to be. For example, in the
extreme case of vanishing dark halo $\gamma=1.9$ works well.

In Section \ref{sec:tori} we explained how to build a model Galaxy using
orbital tori. Torus modelling is best considered an extension of
Schwarzschild modelling, which has long been a standard tool for the
interpretation of data for external galaxies, both in connection with
searches for massive black holes and attempts to understand how early-type
galaxies were assembled. Torus modelling is a more powerful technique
principally because (i) it enables us to quantify  orbits 
by the values taken on them of essentially unique and
physically easily understood isolating integrals, and (ii) it makes it easy to
determine at what velocities a star will pass through any spatial point. We
presented a \df\ of exceptional simplicity, which generates 
a reasonably realistic model of the Galaxy's discs.
 
In Section \ref{sec:comp} we examined in some detail observable quantities in
this model when they are calculated from either the full torus machinery, or
from the \aa. We showed that in the plane, at both $R_0$ and $R=6.5\kpc$, the
distributions of all three components of velocity are reproduced to high
accuracy by the \aa, regardless of whether $\gamma$ is set to zero or unity.
Away from the plane the velocity distribution is sensitive to the weights of
orbits that have relatively large values of $J_z$, with the consequence that
it matters whether the centrifugal potential contains $L_z$ or $|L_z|+\gamma
J_z$, and we find  materially better fits to the distributions of both
$v_\phi$ and $v_z$ when $\gamma=1$ rather than zero.

Regardless of the value of $\gamma$, the \aa\ predicts a value for $\langle v_R\rangle^{1/2}$
that exceeds the true value by an amount that grows with $|z|$, being
$\sim3.4$ per cent at $|z|=1.5\kpc$. The \aa\ yields a value of $\sigma_\phi$
that lies very close to the true value for $|z|<1.3\kpc$, but exceeds the
true value by $\sim10$ percent at $|z|=2\kpc$ because the true value declines
with $|z|$  at $|z|>1.3\kpc$, whereas that obtained from the \aa\ does not. With
$\gamma=1$ the \aa\ yields a value for $\langle v_z\rangle^{1/2}$ that lies within 3 percent of
the true value right up to $|z|=2.3\kpc$.

The \aa\ inevitably predicts that the velocity ellipsoid has two axes parallel
to the plane, so we must turn to the full torus model to discover how the
velocity ellipsoid tilts as one moves away from the plane. We find that its
longest axis points quite close to the Galactic centre. This result emerges
through averaging the quite disparate contributions of individual tori.
Consequently it reflects the structure of the \df\ as much as the
gravitational potential.

From a computational perspective, the \aa\ is extremely convenient, both
because it does not require specialised torus-generating code, and because it
yields $\vJ$ from $(\vx,\vv)$ rather than $(\vx,\vv)$ from $\vJ$.
Consequently, a model's observables can be obtained by integrating over
velocity space, just as traditionally we have obtained the observables of
models with \df s of the form $f(E)$ and $f(E,L)$. While \cite{McMB08} showed
that it is possible to determine $\vJ$ from $(\vx,\vv)$, the procedure used
is iterative and time-consuming, so for this paper observables were estimated
by binning the particles of a discrete realisation obtained by Monte-Carlo
sampling the \df.  Even this procedure is computationally
expensive when enough samples are drawn to make Poisson noise negligible, so
it is very useful to be able to obtain good approximations to $\vJ(\vx,\vv)$
from the \aa, and we anticipate that the \aa\ will be widely used in the
interpretation of observations of the Galaxy.

Paper I and the present paper represent two small steps towards the kind of
Galaxy modelling apparatus that should available before a preliminary Gaia
Catalogue appears in the second half of this decade.  The next big step is to
carry the predictions of models into the space of observables -- such as
apparent magnitudes, parallaxes and proper motions -- and then to explore how
tightly the \df\ of stars can be constrained by data of varying extent and
precision. This step is crucial because distance uncertainties propagate from
observables such as magnitudes and proper motions to correlated errors in
physical quantities such as stellar masses and velocities. We hope to report
on this work soon.

\section*{Appendix A: Evaluating Jacobians}\label{app}

The observables of individual orbits can be obtained from the \df\
$f(\vJ)=\delta(\vJ-\vJ_0)$, where $\vJ_0$ gives the orbits actions. Then
 \begin{eqnarray}\label{eq:moment}
\langle
v_iv_j\rangle_\vx&=&\int\d^3\vv\,v_iv_jf(\vJ)
=\int\d^3\vJ\,\left.{\p(\vv)\over\p(\vJ)}\right|_\vx v_iv_j\delta(\vJ-\vJ_0)\nonumber\\
&=&\sum_i\left(\left.{\p(\vv)\over\p(\vJ)}\right|_\vx v_iv_j\right)_{\vtheta_i},
\end{eqnarray}
 where $\vtheta_i$ are the phases that bring the star to $\vx$. 

Since both the $(\vx,\vv)$ and $(\vtheta,\vJ)$ systems are canonical, there
exists a generating function $G(\vJ,\vx)$ for the transformation
$(\vx,\vv)\leftrightarrow(\vJ,\vtheta)$.
So 
 \[
\theta_i={\p G\over\p J_i}\quad;\quad v_j={\p G\over\p x_j}
\]
 Differentiating again
 \[
\left({\p\theta_i\over\p x_j}\right)_\vJ={\p^2G\over\p x_j\p J_i}=
\left({\p v_j\over\p J_i}\right)_\vx.
\]
 Taking determinants
 \[
\left.{\p(\vtheta)\over\p(\vx)}\right|_\vJ=\left.{\p(\vv)\over\p(\vJ)}\right|_\vx.
\]
 The Jacobian on the left is the orbit's probability density in real space because the
amount of time a star spends in a region of angle space is proportional to
its volume, $\d^3\vtheta$. It makes good physical sense that  the Jacobian in equation (\ref{eq:moment})
is the orbit's probability density.

With toy variables distinguished from true ones by a superscript T, the
generating function of the transformation
$(\vtheta,\vJ)\leftrightarrow(\vtheta^{(\rm T)},\vJ^{(\rm T)})$ between true and toy
angle-action variables is
 \[
S(\vJ,\vtheta^{(\rm T)})=\vJ\cdot\vtheta^{(\rm T)}+\sum_\vn s_\vn(\vJ)\sin(\vn\cdot\vtheta^{(\rm T)}),
\]
 where $\vn$ is a vector with integer components,
 so 
 \begin{eqnarray}\label{eq:giveJtheta}
\vJ^{(\rm T)}&=&\vJ+\sum_\vn s_\vn\vn\cos(\vn\cdot\vtheta^{(\rm T)})\nonumber\\
\vtheta&=&\vtheta^{(\rm T)}+\sum{\p s_\vn\over\p\vJ}\sin(\vn\cdot\vtheta^{(\rm T)}).
\end{eqnarray}
 The torus machine delivers the numerical values of both $s_\vn$ and $\p s_\vn/\p\vJ$.
 We need 
 \[
\left({\p(\vx)\over\p(\vtheta)}\right)_{\vJ}
=\left({\p(\vx)\over\p(\vtheta^{(\rm T)})}\right)_{\vJ}
\left({\p(\vtheta^{(\rm T)})\over\p(\vtheta)}\right)_{\vJ}.
\]
 The inverse of the second Jacobian on the right follows trivially from
equation (\ref{eq:giveJtheta}), and for the
first Jacobian we can write
 \[
\d\vx=\left({\p\vx\over\p\vtheta^{(\rm T)}}\right)_{\vJ^{(\rm T)}}\d\vtheta^{(\rm T)}
+\left({\p\vx\over\p\vJ^{(\rm T)}}\right)_{\vtheta^{(\rm T)}}\d\vJ^{(\rm T)}.
\]
 Dividing through by $\d\vtheta^{(\rm T)}$ and holding $\vJ$ constant we find
 \[
\left({\p(\vx)\over\p(\vtheta^{(\rm T)})}\right)_{\vJ}=\left|
\left({\p\vx\over\p\vtheta^{(\rm T)}}\right)_{\vJ^{(\rm T)}}
+\left({\p\vx\over\p\vJ^{(\rm T)}}\right)_{\vtheta^{(\rm T)}}
\cdot\left({\p\vJ^{(\rm T)}\over\p\vtheta^{(\rm T)}}\right)_{\vJ}\right|.
\]
 The first two matrices on the right involve only toy variables so they are
available analytically, and the third matrix can be obtained from equations
(\ref{eq:giveJtheta}).

\label{lastpage}

\begin{thebibliography}{}

\bibitem[Aumer \& Binney(2009)]{AumerB09}
Aumer M. \& Binney J.J., 2009, MNRAS, 397, 1286

\bibitem[Bahcall \& Soneira(1984)]{BahcallS}
Bahcall J.N., Soneira R.M., 1984, ApJS, 55, 67

\bibitem[Binney(2010)]{B10}
Binney J., 2010, MNRAS, 401, 2318 (Paper I)

\bibitem[Binney \& Spergel(1984)]{BinneyS}
Binney J., Spergel D.N., 1984, MNRAS, 206, 159

\bibitem[Binney \& Tremaine(2008)]{BT08}
Binney J., Tremaine S., 2008, ``Galactic Dynamics'', Princeton University
Press, Princeton (BT08)

\bibitem[Dehnen \& Binney(1998)]{DehnenB97}
Dehnen W., Binney J., 1998, MNRAS, 294, 429

\bibitem[Dehnen \& Gerhard(1993)]{DehnenG}
Dehnen W., Gerhard O.E., 1993, MNRAS, 261, 311

\bibitem[Gebhardt et al.(2003)]{Gebhardt03}
Gebhardt, K. et al (15 authors), 2003, ApJ, 583, 92

\bibitem[Gerhard \& Saha(1991)]{GerhardS}
Gerhard O.E., Saha P., 1991, MNRAS, 251, 449

\bibitem[Gilmore \& Reid(1983)]{GilmoreR}
Gilmore G., Reid N., 1983, MNRAS, 202, 1025

\bibitem[H\"afner et al.(2000)]{Haefner}
Häfner R., Evans N.W., Dehnen W., Binney J., 2000, MNRAS, 314, 433

\bibitem[Kaasalainen(1995)]{Kaasalainen95a}
Kaasalainen M., 1995, MNRAS, 275, 162

\bibitem[Kalnajs(1977)]{Kalnajs} Kalnajs A., 1977, ApJ, 212, 637

\bibitem[Krajnovi\'c et al.(2005)]{Cappellari}
Krajnovi\'c D., Cappellari M., Emsellem E., McDermid R.M.,
de Zeeuw P.T., 2005, MNRAS, 357, 1113

\bibitem[Malin \& Carter(1980)]{MalinC}
Malin D.F., Carter D., 1980, Nature, 285, 643

\bibitem[McMillan \& Binney(2008)]{McMB08}
McMillan, P., Binney J., 2008, MNRAS, 390, 429

\bibitem[Miyamoto \& Nagai(1975)]{MiyamotoN}
Miyamoto M., Nagai R.,~1975, PASJ, 27, 533

\bibitem[Quinn(1984)]{QuinnP}
Quinn P., 1984, ApJ, 279, 596

\bibitem[Robin et al.(2003)]{Robin03}
Robin A.C., Reyl\'e C., Derri\'ere S., Picaud, S., 2003, A\&A, 409, 523

\bibitem[Rowley(1988)]{Rowley88}
Rowley G., 1988, ApJ, 331, 124

\bibitem[Sch\"onrich \& Binney(2009)]{SB09}
Sch\"onrich R., Binney J., 2009, MNRAS, 396, 203

\bibitem[Shu(1969)]{Shu69}
Shu F.H., 1969, ApJ, 158, 505

\bibitem[Skrutskie et al.(2006)]{2MASS}
Skrutskie M.F., et al., 2006, AJ, 131, 1163

\bibitem[Schwarzschild(1979)]{SchwarzI}
Schwarzschild M., 1979, ApJ, 232, 236

\bibitem[Schweizer \& Seitzer(1992)]{SchweizerF}
Schweizer F., Seitzer P., 1992, AJ, 104, 1039

\bibitem[Sharma et al.(2010)]{Sharma10}
Sharma, S., Bland-Hawthorn J., Johnston K.V., Binney J., 2010, ApJ xxx

\bibitem[Siebert et al.(2008)]{Siebert08}
Siebert, A. et al.\ (19 authors) 2008, MNRAS, 391, 793

\bibitem[Smith et al.(2009)]{Smith09}
Smith M.C., Evans N.W., An J.H., 2009, ApJ, 698, 1110

\bibitem[Thomas et al.(2005)]{Thomas05}
Thomas J., Saglia R.P., Bender R., Thomas D., Gebhardt K., Magorrian J., 
Corsini E.M., Wegner G., 2005, MNRAS, 360, 1355

\bibitem[Weinberg(1994)]{Weinberg}
Weinberg, M., 1994, ApJ, 421, 481

\end{thebibliography}
\end{document}